\documentclass[11pt,twoside]{article}
\usepackage{asp2010}

\resetcounters

\begin{document}

\title{MocServer: What \& Where in a few milliseconds.}
\author{Pierre Fernique, Thomas Boch, Ana\"{i}s Oberto, Fran\c{c}ois-Xavier Pineau}
\affil{Observatoire astronomique de Strasbourg, Universit\'{e} de Strasbourg, CNRS, UMR 7550, 11 rue de l'Universit\'{e}, F-67000 Strasbourg, France}

\begin{abstract}
The MocServer is an astronomical service dedicated to the manipulation of data set coverages. This server sets together about 15 000 spatial footprints associated to catalogs, data bases and pixel surveys. Thanks to the Multi-Order Coverage map coding method (MOC1.0 IVOA standard), the MocServer is able to provide in a few milliseconds the list of data set identifiers intersecting any polygon on the sky. The MOC server has been deployed in June 2015 by the Centre de Donnees astronomiques de Strasbourg. It is operational and already in use by Aladin Desktop and Aladin Lite prototype versions.
\end{abstract}

\section{Context and motivations}

The interoperability in astronomy discipline are regularly facing to this basic question: {\em Which data collections, with such or such characteristics, are available in a specific sky region?} This question is the basis of a large collection of use cases. We illustrate them with these three examples:
\begin{itemize}
  \item Which reference surveys are covering my M33 observation field? Notably those after May 2010?
	\item Which catalogs have sources simultaneously in HST and GALEX observations, and for which there are velocity measurements?
	\item From my object collection, which of them are not in Simbad nor in NED yet?
\end{itemize}

Several technical solutions are already in use, generally based on classical data bases associated to interoperability protocols such as IVOA TAP \citep{TAP}. But the capabilities and performances, notably spatially, are generally poor and, concretely, do not allow to solve rapidly and easily these kind of uses cases.

The solution that we present in this paper is simple, global and extremely fast. All these use cases are solved in a few milliseconds. It is based on Multi-Order Coverage maps, also named MOC.

\section{What is a MOC?}
A MOC is a simple and efficient method for describing a sky region. Its principle is based on the list of the HEALPix cell indices covering a specific region, grouped hierarchically. Its basic goal is to compare sky regions as fast as possible. It is described in \citep{2012ASPC..461..347F}.

MOC is a standard recommended by IVOA since June 2014 \citep{MOC}.

\section{Why a MOC server?}

A collection of MOCs manages together by a dedicated server allows one to retrieve, not only the coverage of each data set, but also the list of data set identifiers matching a specific sky region.

The algorithm is based on a simple loop checking for each MOC if the intersection with the user sky region MOC is empty or not. If the intersection is not empty, the identifier associated to the MOC is added to the resulting list.

\section{MOC server property extension}

The MocServer is designed to manage coverage maps associated to data set identifiers. However, by extending the MocServer content with some properties associated to each data set, for instance the title, the description, some keywords describing data set characteristics, this server becomes a fast and powerful meta-data server, spatially indexed.

\section{How it is implemented ?}

The MocServer, implemented at CDS in June 2015, is basically a 3 000 java line code Tomcat servlet. It manages thousands of MOCs and properties, harvested regularly by scripts from Simbad data base and VizieR catalog service and other partner data bases.

The MocServer is searchable by regions: circle, polygon or MOC, and/or by keywords via a simple HTTP GET or POST method. It provides the identifier list and/or the property records of the data sets found in the region. It is also able to build on the fly the union or the intersection of the resulting data set MOCs.
Several output formats are supported: simple ASCII and JSON for identifiers lists and records lists, JSON and FITS for MOC results. The data set identifiers are specified as IVORNs following the IVOA recommendation.

\section{Performances and examples}

Only 400 MB are required for manipulating the 15 000 MOCs and associated properties from VizieR and Simbad. This characteristic allows to build a server managing directly the data in memory. This simple and robust solution provides very fast responses. Some examples are provided below. The network transfer time is not taken into account in the response times.

\begin{itemize}
\item IDs of all data sets in 5 deg around M31:\\
http://...$?$ RA=10.67305 \& DEC=41.26875 \& SR=5 \\
=$>$ 5ms
\item IDs of HST collections overlapping SDSS observations:\\
http://...$?$ ivorn=*HST* \& url=http://urlMocSDSS \\
=$>$ 21 ms
\item MOC union of all Seyfert data sets \\
http://...$?$ obs\_astronomy\_kw=Seyfert* \& get=moc \\
=$>$ 44ms
\item MOC union of all A\&AS tables \\
http://...$?$ ivorn=CDS/J/A+AS/* \& get=moc \\
=$>$ 492ms
\end{itemize}

\articlefigure{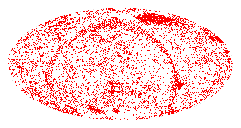}{Fig1}{Coverage of all A\&AS published tables displayed as a global MOC union}

For increasing the performance of the MocServer, internally, the MOCs are sorted by size. Due to the nature of MOC, its size in memory is directly in relation with the distribution of the data on the sky \citep[cf. ][]{2015A+A...580A.132R}. In other terms, more the MOC is large, more its data is fragmented on the sky. The loop for comparing user region with MOCs must start from the smallest MOCs to the largest in order to remove as fast as possible false candidates. 

Based on the same MOC characteristic, the computation of the intersection of several MOCs is faster by starting from the smallest one. Conversely, the computation of the unions of MOCS will be faster by starting from the largest one.

A stress test has demonstrated that the MocServer is fully supporting several millions queries per day with the current implementation.

\section{For whom tools ?}

The MocServer deployed by CDS is dedicated to any remote clients, VO oriented or not. Its speed allows to implement dynamic interfaces. It is already in use by Aladin Desktop \citep{Aladin} and Aladin Lite \citep{AladinLite}. These clients are continuously querying the MocServer with the sky region covering the user current view to update the data set list concerned by this current user view.

The service entry point and documentation are available at this following address: http://alasky.unitra.fr/MocServer/query

\articlefigure[scale=0.3]{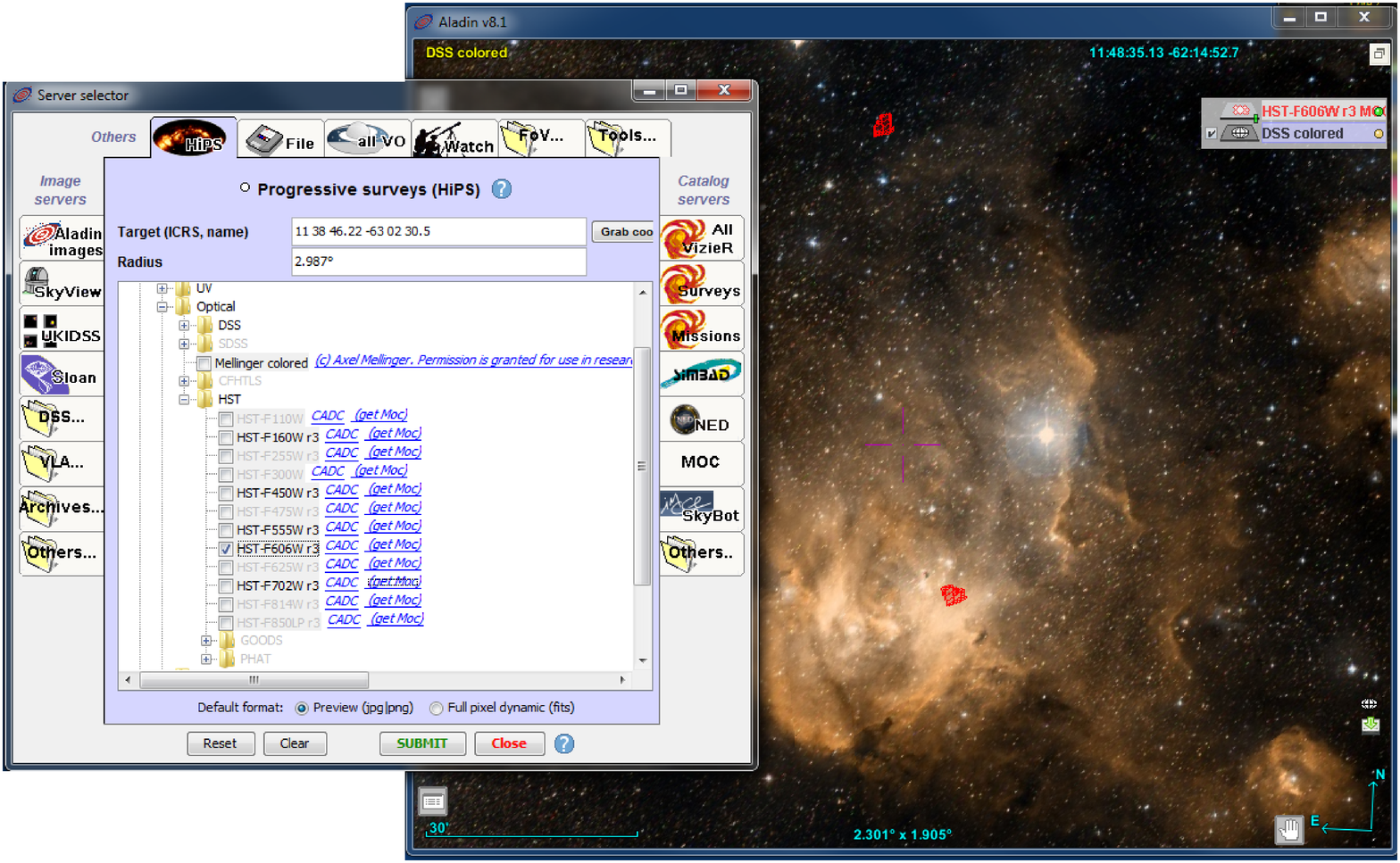}{Fig2}{Aladin Desktop dynamic HiPS tree - Only black items in the tree have observations in the current user view.}

\bibliography{P037}

\end{document}